\RequirePackage{amsmath}
\documentclass[12pt]{iopart}
\pdfoutput=1

\usepackage{graphicx}
\usepackage[utf8]{inputenc}

\usepackage{amsmath,amsfonts,amssymb}

\usepackage{xcolor}
\colorlet{mylinkcolor}{blue!66!black!80}
\usepackage[colorlinks=true,linkcolor=mylinkcolor,citecolor=mylinkcolor,filecolor=cyan,urlcolor=mylinkcolor,breaklinks=true]{hyperref}
\usepackage{cite}

\newcommand{\avg}[1]{\langle#1\rangle}

\newcommand{\pmax}{p^{\rm max}}
\newcommand{\Pmax}{P^{\rm max}}
\newcommand{\pmin}{p^{\rm min}}
\newcommand{\Pmin}{P^{\rm min}}

\newcommand{\dd}{{\rm d}}
\newcommand{\Prob}{{\rm Prob}}
\newcommand{\Surv}{\mathcal{P}}
\newcommand{\pmaxLD}{p^{\rm max}_{\rm LD}}
\newcommand{\pminLD}{p^{\rm min}_{\rm LD}}
\newcommand{\Peq}{P^{\rm eq}}

\newcommand{\Pmaxmin}{P^{\kappa}}
\newcommand{\pmaxmin}{p^{\kappa}}
\newcommand{\pmaxminLD}{p^{\kappa}_{\rm LD}}
\newcommand{\maxmin}{\kappa}

 \usepackage{amsmath}
 
\begin{document}
 \title[Large deviations of extremes from first passage times]{Extreme value statistics of ergodic Markov processes from first passage times in the large deviation limit} 
 \author{David Hartich and Alja\v{z} Godec}
\address{Mathematical Biophysics Group, Max-Planck-Institute for
  Biophysical Chemistry, G\"{o}ttingen 37077, Germany}
\eads{\href{mailto:david.hartich@mpibpc.mpg.de}{david.hartich@mpibpc.mpg.de},\href{mailto:agodec@mpibpc.mpg.de}{agodec@mpibpc.mpg.de}}

\begin{abstract}
  Extreme value functionals of stochastic processes are inverse functionals of the first passage time --  a connection
  that renders their probability distribution functions
  equivalent. Here, we deepen this link and establish a framework for analyzing
  extreme value statistics of ergodic reversible Markov processes in
  confining potentials on the hand of the underlying relaxation
  eigenspectra. We derive a chain of inequalities, which bounds the
  long-time asymptotics of first passage densities, and thereby extrema, from above and from
  below. The bounds involve a time integral of the transition
  probability density describing the relaxation towards
  equilibrium. We apply our general results to the analysis of extreme value
  statistics at long times in the case of Ornstein-Uhlenbeck process
  and a 3-dimensional Brownian motion confined to a sphere, also known as Bessel process.  
We find that even on time-scales that are shorter than the equilibration time,
the large deviation limit characterizing long-time asymptotics can approximate the statistics of extreme values remarkably well.
Our findings provide a novel perspective on the
study of extrema beyond the established limit theorems for
sequences of independent random variables and for asymmetric diffusion processes beyond a constant drift.
\end{abstract}


\date{\today}
\section{Introduction}
The statistical properties of extreme values, which correspond to
record-breaking events of a stochastic process, attracted increasing interest in various fields of research over the past decades.
For example, climate changes were found to be reflected in the appearance of extreme (record-breaking) temperatures
\cite{redn06,werg10}, rainfall \cite{nada05,maju18arxiv}, and possibly other extreme weather conditions \cite{coum12}.
Statistics of records are also important in the context of earthquakes \cite{sorn96},
in studies of stock pricing in economics \cite{embr97,sabi14}, sports \cite{gemb02,ben-07}, and in the theory of random matrices \cite{dean06,dean08,elia18arxiv} to name but a few (see, e.g., Ref.~\cite{werg13,godr17} for a more detailed overview).

In sequences of independent random variables extreme values approach
one of the three classes of limiting  distributions, which are denoted
by the Gumbel \cite{gumb58}, Fr\'echet, and Weibull distributions
(see, e.g., Refs.~\cite{bert06,sabh07,krug07,luca12,werg13}). However, as soon as consecutive time steps of a stochastic process become correlated, a theoretical
discussion of the statistics of extrema becomes more challenging
\cite{comt05}. In this case universal laws have been discovered, for example, for processes with symmetric step-length distributions
\cite{maju08,maju10}. Subsequent studies also investigated
correlations between records \cite{beni16,beni16a} as well as
their persistence \cite{godr09,maju12,szab16} and number
\cite{eder13}, and extensions have been made to processes with
constant drift \cite{werg11,maju12,mart18,moun18} (see also
Ref.~\cite{kind17} for an interesting experiment with trapped Cs
atoms). A recent physical application includes the observation
that the mean value of the minimum of the entropy production in stationary driven systems is bounded by the negative value of Boltzmann's constant  ``$-k_{\rm B}$''  \cite{neri17}, which is also confirmed by experiments \cite{sing17arxiv}.

More broadly, a deep and important connection has been established, relating the statistics of extreme values to
first passage times \cite{maju10,bray13,werg13,sche14}.
In this work we deepen this connection between the first passage and the extremum functional, which allows us
to obtain the statistics of extreme values in finite time for Markovian diffusion processes in
confining potentials on time-scales, where consecutive time-steps remain correlated. 
Exploiting further a duality between first passage processes and ensemble propagation
\cite{hart18,hart18a_arxiv}
we derive a chain of inequalities, which bound the long time
asymptotics (i.e., the large deviation limit) of the probability densities of extrema
both from above and from below. As we will show, the large deviation limit
approximates the probability density of extreme values surprisingly well even on relatively short time-scales.

The paper is organized as follows. In Sec.~\ref{sec:fundamentals} we recapitulate the well-known connection between
distributions of extrema and first passage time densities. We then utilize recent findings on the large time asymptotics
of first passage time densities \cite{hart18,hart18a_arxiv} to
determine distributions of extrema in the large deviation limit. The
usefulness of our general results is demonstrated in Sec.~\ref{sec:examples},
by determining the long-time statistics of maxima of the
Ornstein-Uhlenbeck process and the statistics of the minimum of the
3-dimensional Brownian motion (Bessel process) confined to a
sphere. All analytical results are corroborated by Brownian dynamics
simulations. We conclude in Sec.~\ref{sec:conclusion}.

\section{Fundamentals}
\label{sec:fundamentals}
\subsection{Extreme values from first passage times}
\label{sec:max_vs_FP}
We consider processes governed by an overdamped Langevin equation
\begin{equation}
 \dot x_t=-\beta D U'(x_t)+\xi_t
 \label{eq:Langevin}
\end{equation}
where $U'(x)=\partial_x U(x)$ is the gradient of a potential $U(x)$ and $\xi_t$
stands for  Gaussian white noise with zero mean and covariance
$\avg{\xi_t\xi_{t'}}=2D\delta(t-t')$. Without any loss of generality we set the inverse temperature $\beta$ and diffusion coefficient $D$ to unity ($\beta\equiv D\equiv1$), i.e., free energies $U$ are expressed in units of $k_{\rm B} T$.
The Fokker-Planck equation corresponding to the Langevin equation~\eqref{eq:Langevin}
reads \cite{gard04}
\begin{equation}
\partial_t P(x,t|x_0)=\partial_x \big[\partial_x+U'(x)\big]P(x,t|x_0)\equiv \mathcal{L}_{\rm FP}P(x,t|x_0),
 \label{eq:FPE}
\end{equation}
where $P(x,t|x_0)=\avg{\delta(x-x_t)}$ is the normalized probability
density for a particle starting from $x_0$ to be found at position $x_t=x$ at time $t$
with the initial condition $P(x,0|x_0)=\delta(x-x_0)$.
The probability density function relaxes to
the normalized Boltzmann-Gibbs equilibrium density $\Peq(x)\equiv P(x,\infty|x_0)\propto \e^{-U(x)}$
for any $x_0$, which requires a sufficiently confining potential $U(x)$.

\begin{figure}
\includegraphics{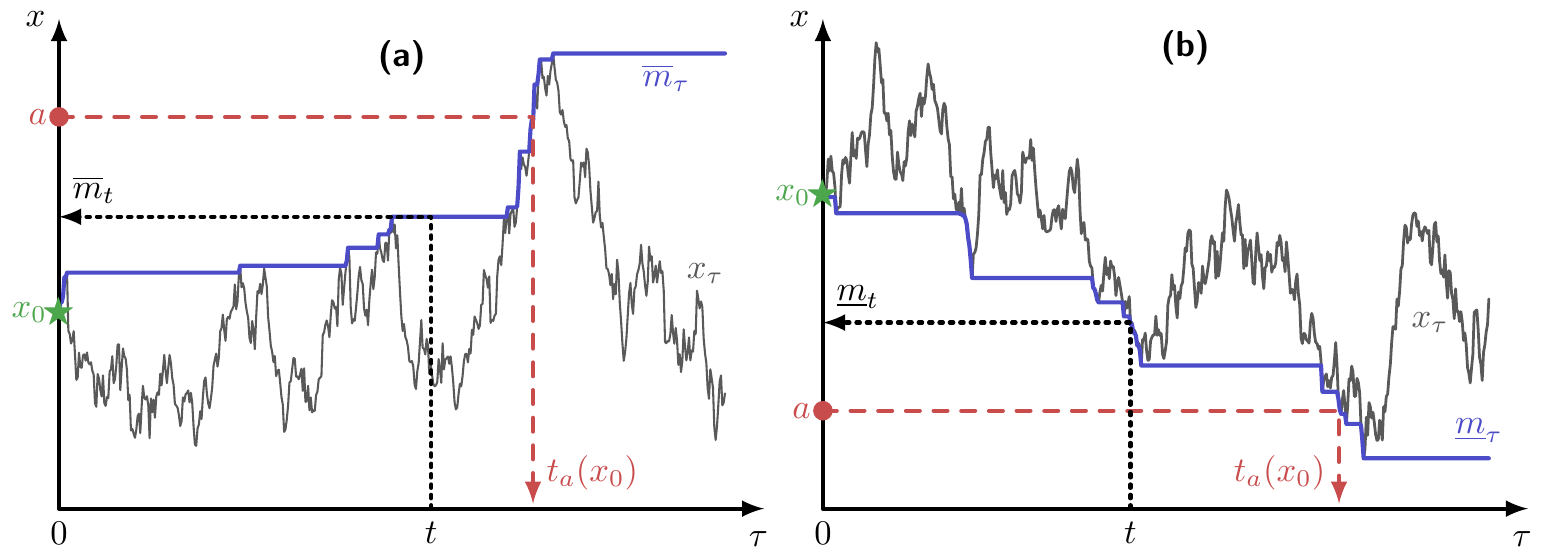}
 \caption{Schematics of the extreme value functional versus first passage time functional for a process $\{x_\tau\}_{0\le\tau}$. (a)~Schematics of the maximum functional $\overline{m}_\tau$ (thick blue line) of the process $x_\tau$ (thin gray line) as function of time $\tau$;
 the dotted black arrow indicates the functional of the maximum $\overline{m}_t$ and the dashed red arrow represents first passage time $t_a(x_0)$;
 the arrows indicate the equivalence between $a> \overline{m}_t$ and
 $t_a(x_0)> t$. (b)~The minimum functional $\underline{m}_\tau$ (thick
 blue line) defined analogously to~(a), whereas $a<\underline{m}_t$ is
 here equivalent to $t_a(x_0)> t$.}%
 \label{fig:min_max}%
\end{figure}

While the probability density $P(x,t|x_0)$ only depends on the initial and final states,
the extreme values are functionals that depend on the entire history along a trajectory $\{x_\tau\}_{0\le\tau \le t}$. We define
the maximum and the minimum of the process $x_t$
as
\begin{equation}
 \overline{m}_t\equiv\max_{0\le \tau\le t}(x_\tau) \qquad\text{and}\qquad
  \underline{m}_t\equiv\min_{0\le \tau\le t}(x_\tau),
\end{equation}
respectively. For a given initial condition $x_0$
the extrema satisfy $\overline{m}_t\ge x_0$ as well as $\underline{m}_t\le x_0$,
where $\overline{m}_t$ is non-decreasing and $\underline{m}_t$ non-increasing in time $t$ (see Fig.~\ref{fig:min_max}). 
It can be shown that the first passage time defined as
\begin{equation}
 t_a(x_0)\equiv\min_{\tau\ge 0}(\tau|x_\tau=a)
\end{equation}
is an inverse functional of the extrema.
To see that we consider the distribution function of the maximum of the process $\overline m_t=\max_{0\le \tau\le t}(x_\tau)$, i.e., the probability that $\overline{m}_t$ exceeds the value $a\ge x_0$, $\Pmax(a|t, x_0)$, which satisfies
\begin{equation}
 \Pmax(a|t, x_0)\equiv\Prob\left[a>\overline{m}_t\right]=\Prob\left[t_a(x_0)> t\right]\equiv  \Surv_a(t|x_0),
\label{eq:max_vs_FP}
 \end{equation}
 where $\Surv_a(t|x_0)$ is called the survival probability (see also Fig.~\ref{fig:min_max}a).
Eq.~\eqref{eq:max_vs_FP} can be interpreted as follows: each path, whose maximum $\overline m_t$ after time $t$ is smaller
than $a$, must have a first passage time from $x_0$ to $a\ge x_0$, $t_a(x_0)$,
larger than $t$. Eq.~\eqref{eq:max_vs_FP} connects the first passage
time functional $t_a(x_0)$ (where time is stochastic and position is
fixed to $a$) to the maximum value functional $\overline{m}_t$ (where
time is fixed to $t$ and the position is stochastic). Note that the survival probability can also be expressed as the integral
over the first passage time density $\wp_a$ via
\begin{equation}
 \Surv_a(t|x_0)=\int_t^\infty\wp_a(\tau|x_0)\dd \tau,
 \label{eq:Surv_int}
\end{equation}
i.e., $\wp_a(\tau|x_0)=-\partial_t\Surv_a(t|x_0)$.

The minimum of a process $\underline{m}_t=\min_{0\le \tau\le t} x_t$
can be studied in a similar manner as the maximum, since
$\underline{m}_t$ is equivalent to the maximum of the reflected process $-x_\tau$, i.e.,  $\underline{m}_t=-\max_{0\le \tau\le t}(-x_\tau)$.
Hence, in the case of the minimum  ($a\le x_0$) Eq.~\eqref{eq:max_vs_FP}
holds with the replacement $ \Pmin(a|t, x_0)=\Prob\left[a<\underline{m}_t\right]=\Surv_a(t|x_0)$, which is illustrated in Fig.~\ref{fig:min_max}b.
For convenience, we simply refer to $\Pmaxmin$ as the extremum distribution function, which in the case $x_0\le a$ corresponds to the maximum distribution ($\maxmin={\rm max}$) and for $x_0\ge a$ to the minimum distribution ($\maxmin={\rm min}$).
The density of the extremum ($\maxmin={\rm max},{\rm min}$) in either case is then given by the slope of the distribution function
\begin{equation}
\pmaxmin(a|t, x_0)=\big|\partial_a\Pmaxmin(a|t,x_0)\big|=\big|\partial_a\Surv_a(t|x_0)\big|,
\label{eq:pmax_slope}
\end{equation}
where in the second step we used Eq.~\eqref{eq:max_vs_FP}.
Eq.~\eqref{eq:pmax_slope} describes the density of maxima ($\maxmin={\rm max}$) for  $a\ge x_0$
and the density of the minima ($\maxmin={\rm min}$) for 
$a\le x_0$.
For example, the mean value of the maximum and minimum are given, respectively, by
\begin{equation}
 \begin{aligned}
  \avg{\overline{m}_t}&=\int_{x_0}^\infty \pmax(a|t,x_0)a\dd a=x_0+\int_{x_0}^\infty\Surv_a(t|x_0)\dd a
  ,\\
  \avg{\underline{m}_t}&=\int_{-\infty}^{x_0} \pmin(a|t,x_0)a\dd a=x_0-\int_{-\infty}^{x_0}\Surv_a(t|x_0)\dd a,
 \end{aligned}
\end{equation}
where we used Eq.~\eqref{eq:pmax_slope} and performed a partial integration in the last step in both lines.

In the following subsection we focus on the probability density function of the two extrema $\pmax$ and $\pmin$, whereas further discussions on the mean of extreme value fluctuations ($\avg{\underline{m}_t}$ or $\avg{\overline{m}_t}$) 
can be found, for example, in Refs.~\cite{coff98,comt05,maju10,beni16,beni16a}. 
Notably, $t_a(x_0)$ -- the  first passage time from $x_0$
to $a\ge x_0$ -- is unaffected by the potential landscape $U(x)$ beyond
$x\ge a$, which according to Eq.~\eqref{eq:max_vs_FP} implies that any
two potentials $U_1(x),U_2(x)$ with $U_1(x)=U_2(x)$ for all $x\le R$ generate the same maximum distribution $\Pmax$ for all $ a\le R$.

\subsection{First passage time statistics from ensemble propagation}
According to Eq.~\eqref{eq:max_vs_FP} the problem of determining the statistics of the extremum $\Pmax(a|t,x_0)$ 
is in fact equivalent to determining the survival probability $\mathcal{P}_a(t|x_0)$, or, according to Eq.~\eqref{eq:Surv_int}, to determining the first passage time density $\wp_a(t|x_0)=-\partial_t \Surv_a(t|x_0)$, which will be the central goal of this section.

We determine the first passage time density (or survival probability)
using the renewal theorem \cite{sieg51}
\begin{equation}
 P(a,t|x_0)=\int_0^tP(a,t-\tau|a)\wp_a(\tau|x_0)\dd \tau,
 \label{eq:renewal}
\end{equation}
reflecting the fact that all the paths starting from $x_0$
and ending up in $a$ after time $t$ by construction must reach
$a$ for the first time at some time $\tau\le t$, and then return to $a$ again after time $t-\tau$.
We have recently established a duality between first passage and relaxation processes, i.e., between $P$ and $\wp_a$, which will allow us to solve Eq.~\eqref{eq:renewal} for $\wp_a$ in the following manner (see also Refs.~\cite{hart18, hart18a_arxiv} for more details).

First, we Laplace-transform\footnote{%
The Laplace transform of a function $f(t)$ is defined by $\tilde f(s)=\int_0^\infty f(t)\e^{-st}\dd t$.}
the renewal theorem \eqref{eq:renewal} in time $(t\to s)$,
which converts the convolution to a product, $\tilde P(a,s|x_0)=\tilde P(a,s|a)\tilde\wp_a(s|x_0)$, implying
\begin{equation}
 \tilde\wp_a(s|x_0)=\frac{\tilde P(a,s|x_0)}{\tilde P(a,s|a)},
 \label{eq:renewal_Laplace}
\end{equation}
where $\tilde\wp_a(s|x_0)$ is the Laplace transform of the first passage time density
and $\tilde P(x,s|x_0)$ obeys the Laplace transformed Fokker-Planck equation \eqref{eq:FPE}
\begin{equation}
 [\mathcal{L}_{\rm FP}-s]\tilde P(x,s|x_0)=-\delta(x-x_0)
\end{equation}
with natural boundary conditions.
The next step is to render Eq.~\eqref{eq:renewal_Laplace} explicit in the time domain, i.e.,
to find the explicit inverse Laplace transform $\tilde\wp_a(s|x_0)\to \wp_a(t|x_0)$.

Therefore, recalling that we consider sufficiently confining potentials $U(x)$,
there exist a spectral expansion of the Fokker-Planck operator
$\mathcal{L}_{\rm FP}$ with discrete eigenvalues $-\lambda_k\le 0$ for
$k=0,1,\ldots$ and corresponding symmetrized nontrivial solutions $\psi_k(x)$ to
the eigenequation $\mathcal{L}_{\rm FP}\psi_k(x)=-\lambda_k\psi_k(x)$, which are assumed to be normalized $\int\psi_k(x)^2/\Peq(x)\dd x \equiv 1$.
The ground state $\psi_0$ corresponding to eigenvalue $\lambda_0=0$
represents the equilibrium Boltzmann distribution $\Peq(x)=\psi_0(x)$.
Using the eigenfunctions $\{\psi_k\}$ and eigenvalues $\{\lambda_k\}$
defined this way, the Laplace transform of the ensemble propagator can be written in the form
\begin{equation}
 \tilde P(a,s|x_0)=\frac{\Peq(a)}{s}+\sum_{k=1}^\infty\frac{\psi_k(a)\psi_k(x_0)/\Peq(x_0)}{s+\lambda_k},
 \label{eq:P(a,s|x_0)}
\end{equation}
where $\psi_k(x)/\Peq(x)\equiv\psi_k^\dagger(x)$
are in fact the eigenfunctions to the adjoint of $\mathcal{L}_{\rm FP}$, that is
$\mathcal{L}_{\rm
  FP}^\dagger\psi^\dagger(x)=-\lambda_k\psi^\dagger(x)$. Note that in
our previous work we used the equivalent non-symmetric eigenspectrum with right and left
eigenfunctions $\psi_k$ and $\psi_k^\dagger$, respectively \cite{hart18,hart18a_arxiv,lapo18}.
Since the Laplace transform of a function $f$ with a simple pole $\tilde f(s)=(s+\lambda)^{-1}$ yields in the time domain an exponentially decaying function $f(t)=\e^{-\lambda t}$ with rate $\lambda$,
we can interpret the eigenvalues $\lambda_k$ as relaxation rates, which characterize the speed at which the dynamics governed by Eq.~\eqref{eq:FPE} approaches the equilibrium $\Peq(x)\propto\e^{-U(x)}$.

The Laplace transform of the first passage time density $\tilde \wp_a(s|x_0)$, as well
has simple poles, which are located at $s=-\mu_k$ ($k=1,2,\ldots$) and need to be determined for Eq.~\eqref{eq:renewal_Laplace} to be written as \cite{hart18}
\begin{equation}
\tilde \wp_a(s|x_0)=\sum_{k=1}^\infty\frac{w_k(a,x_0)\mu_k(a)}{\mu_k+s}.
\label{eq:FP_Laplace}
\end{equation}
The expansion in Eq.~\eqref{eq:FP_Laplace} can formally be found by determining the zeros  $s=-\mu_k$ that solve $\tilde P(a,s|a)=0$,
to which we refer as first passage rates $\mu_k$.
Determining all first passage rates $\mu_k$, while doable in general, is rather involved
and is described in \cite{hart18,hart18a_arxiv}, whereas detailed information on
the determination of slowest rate $\mu_1$, to which we refer to as large deviation limit,
can be found in Sec.~\ref{sec:LD} below as well as in \cite{gode16}. If all first passage rates $\{\mu_k\}$ are known, we can obtain
the corresponding weights $w_k(a,x_0)$ in Eq.~\eqref{eq:FP_Laplace} directly from Eq.~\eqref{eq:renewal_Laplace}
using Cauchy's residue theorem
\begin{equation}
 w_k(a,x_0)\mu_k(a)=\frac{\tilde P(a,s|x_0)}{\partial_s\tilde P(a,s|a)}\bigg|_{s=-\mu_k}.
 \label{eq:residue}
\end{equation}
Dividing Eq.~\eqref{eq:residue} by $\mu_k(a)$ yields the ``weights'' $w_k(a,x_0)$, which according 
to Eq.~\eqref{eq:FP_Laplace} are normalized such that $\tilde \wp_a(0|x_0)=\sum_kw_k(a,x_0)=1$.
Eq.~\eqref{eq:FP_Laplace} in turn immediately yields the first passage time density
\begin{equation}
 \wp_a(t|x_0)=\sum_{k>0}w_k(a,x_0)\mu_k(a)\e^{-\mu_k(a)t},
 \label{eq:FPTD}
\end{equation}
and the corresponding survival probability
\begin{equation}
 \Surv_a(t|x_0)= \int_t^\infty\wp_a(\tau|x_0)\dd \tau=\sum_{k>0}w_k(a,x_0)\e^{-\mu_k(a)t},
  \label{eq:FPTDint}
\end{equation}
where we used Eqs.~\eqref{eq:Surv_int} and \eqref{eq:FPTD}.
Inserting Eq.~\eqref{eq:FPTDint} into \eqref{eq:pmax_slope} allows us
to rewrite the probability density of the extremum to have value $a$
at time $t$ as
 \begin{align}
 \pmaxmin(a|t,x_0)&=\big|\partial_a \Pmaxmin(a|t, x_0)\big| =\big|\partial_a\Surv_a(t|x_0)\big|
 =\bigg|\sum_{k>0}\partial_a \big[w_k(a,x_0)\e^{-\mu_k(a)t}\big]\bigg|,
 \label{eq:pmax}
\end{align}
where $\maxmin={\rm max}$ or $\maxmin={\rm min}$. 

\subsection{Large deviation limit}
\label{sec:LD}
At long times the extremum ($\overline{m}_t$ or $\underline{m}_t$)
will be dominated by extreme fluctuations of the process $x_t$ that
are not reflected by the ``typical'' equilibrium measure given by
$P^{\rm eq}(x)\propto\e^{-U(x)}$. As a result, the extreme value
distribution may differ substantially from the
equilibrium Boltzmann distribution $\Peq(x)$. Fortunately, at long
times the first passage distribution will be dominated solely by the slowest
first passage time-scale $1/\mu_1(a)$,
which leads to what we refer here to as the \emph{large deviation limit} that reads
\begin{equation}
\pmaxminLD(a|t,x_0)\equiv \big| \partial_a w_1(a,x_0)\e^{-\mu_1(a)t}\big|\simeq \pmaxmin(a|t,x_0),
\label{eq:LDdef}
\end{equation}
where ``$\simeq$'' denotes the asymptotic equality in the limit
$t\to\infty$ and $\maxmin={\rm max},{\rm min}$. An explicit general
method to determine $w_1(a,x_0)$ and $\mu_1(a)$ can be found in
\cite{gode16,hart18,hart18a_arxiv}. We note that the large deviation limit becomes exact in the long time limit
$\e^{-\mu_1(a)t}\gg \e^{-\mu_2(a)t}$ as well as whenever $w_1(a,x_0)\gg |w_{k\ge2}(a,x_0)|$ holds.

We recall that according to Eq.~\eqref{eq:FP_Laplace} each first passage rate, $\mu_k$,
is located at a simple pole (at $s=-\mu_k$)  of $\tilde\wp_a(s|x_0)$, which according to Eq.~\eqref{eq:renewal_Laplace}, is also a root of $\tilde P(a,s|a)$.
Hence, the large deviation limit ``$\mu_1$'' is characterized by the root ($s<0$) closest to the origin, $s=-\mu_1$, solving $\tilde P(a,s|a)=0$. In order to determine $\mu_1$ exactly, we Taylor-expand the function
\begin{equation}
 f(s)= s\tilde P(a,s|a)=\sum_{n\ge0}\sigma_n s^n/n!
 \label{eq:Taylor_series}
\end{equation}
around $s=0$, where $\sigma_n$
is the $n$th derivative of $f$ with respect to $s$, which according to Eq.~\eqref{eq:P(a,s|x_0)}
holds for all $s$ within the radius of convergence $|s|<\lambda_1$.%
\footnote{The radius of convergence is limited by the pole of $f(s)=s\tilde P(a,s|a)$ which is closest to the origin. According to Eq.~\eqref{eq:P(a,s|x_0)} the closest pole to $s=0$ is located at $s=-\lambda_1$, yielding a converging sum Eq.~\eqref{eq:Taylor_series} for all $|s|<\lambda_1$.}
Note that $f(s)=s\tilde P(a,s|a)$ has the same non-trivial roots but, in contrast to $\tilde P(a,s|a)$, does not have a pole at the origin (see Eq.~\eqref{eq:P(a,s|x_0)}), which is why we are always allowed to expand $f$ as in Eq.~\eqref{eq:Taylor_series}.
The closest non-trivial zero $s=-\mu_1$ with $0<\mu_1\le\lambda_1$ can then be formally be found by a Newton's iteration, which in terms of a series of almost triangular matrices reads explicitly \cite{hart18a_arxiv} (see also \cite{gode16,hart18})
\begin{equation}
\mu_1(a)=\sum_{n=1}^\infty\frac{\sigma_0^n}{\sigma_1^{2n-1}}\frac{\det\boldsymbol{\mathcal{A}}_n}{(n-1)}
\label{eq:mu1_Newton}
\end{equation}
where $\boldsymbol{\mathcal{A}}_n$ is a $(n-1)\times(n-1)$ almost triangular matrix with elements ($i,j=1,\ldots, n-1$)
\begin{equation}
\mathcal{A}_n^{ij}\equiv \frac{\sigma_{i-j+2}\Theta(i-j+1)}{(i-j+2)!}
 \times\left\{\begin{array}{@{}ll@{}}
 i&\quad\text{if $j=1$,}\\
 n(i-j+1)+j-1&\quad\text{if $j>1$,}
 \label{eq:A}
\end{array}
\right.
\end{equation}
with $\Theta(l)=1$ if $l\ge0$ and $\Theta(l)=0$ if $l<0$ as well as $\det\boldsymbol{\mathcal{A}}_1\equiv1$.
Eq.~\eqref{eq:mu1_Newton} exactly determines the first non-trivial root, $f(s)=0$ with $s=-\mu_1$, at which the right hand side of Eq.~\eqref{eq:Taylor_series} vanishes.
It should be noted that determining $\mu_1$ in Eq.~\eqref{eq:mu1_Newton} requires 
only $\tilde P(a,s|a)$ or the coefficients $\sigma_n$ from Eq.~\eqref{eq:Taylor_series}, whereas the expansion Eq.~\eqref{eq:P(a,s|x_0)} including the eigenvalues $\{\lambda_k\}$ is generally not required to be known.
The weight $w_1(a)$ can then be deduced from Cauchy's residue theorem Eq.~\eqref{eq:residue}
\begin{equation}
 w_1(a)=\frac{\tilde P(a,-\mu_1(a)|x_0)}{\mu_1(a) \partial_s\tilde P(a,s|a)}\bigg|_{s=-\mu_1(a)}.
 \label{eq:w1}
\end{equation}
Equation \eqref{eq:LDdef} with Eqs. \eqref{eq:mu1_Newton} and \eqref{eq:w1} fully characterize the large deviation limit of the density of the extreme value $\pmaxLD(a|x_0)$.

\subsection{Large deviation limit in the presence of a spectral gap}
In the large time limit the probability mass of the extremum $\pmax(a|t,x_0)$ or
$\pmin(a|t,x_0)$ concentrates at the potential boundaries (i.e.,
$U(a)\gg k_{\rm B}T$), such that we can accurately approximate $\mu_1$
by truncating Eq.~\eqref{eq:mu1_Newton} already after the first term yielding (see Ref.~\cite{hart18} for more details) 
\begin{equation}
\tilde\mu_1(a) \equiv\frac{\sigma_0}{\sigma_1}\approx \mu_1(a),
 \label{eq:LD_approx}
\end{equation}
where using Eq.~\eqref{eq:P(a,s|x_0)} we can identify
\begin{equation}
 \sigma_0=\Peq(a)\qquad \text{and}\qquad \sigma_1=\int_0^\infty [P(a,t|a)-\Peq(a)]\dd t.
 \label{eq:LD_approx_int}
\end{equation}
Since $f(s)\equiv s\tilde P(a,s|x_0)=\sigma_0+\sigma_1
s+\mathrm{O}(s)^2$, we expect Eq.~\eqref{eq:LD_approx} to be quite
accurate as soon as the formal condition $\tilde\mu_1\ll\lambda_1$ is
met, where $\lambda_1$ from Eq.~\eqref{eq:P(a,s|x_0)} is the slowest rate at which the system approaches the equilibrium density \cite{hart18}.
Note that $\lambda_1$ in fact does not need to be known, as
Eq.~\eqref{eq:LD_approx} necessarily becomes accurate at sufficiently
high potential values $U(a)$, such that $a$ is not located in the
deepest point in the potential \cite{hart18}. This also follows from
the work of Matkowsky and Schuss, who have shown that $\lambda_1$ is the expected time to overcome the barriers on the way to the deepest potential well \cite{matk81}.

In fact, at very long times $t\to\infty$ the probability mass
$\pmaxmin(a|t,x_0)\simeq\pmaxminLD(a|t,x_0)$ (with $\maxmin={\rm
  \max},{\rm min}$) will inevitably be pushed towards the boundaries
with high potential values, which will again render Eq.~\eqref{eq:LD_approx} asymptotically
exact in the limit $U(a)\to\infty$.
To prove that Eq.~\eqref{eq:LD_approx} indeed becomes asymptotically exact, we inspect Eq.~\eqref{eq:P(a,s|x_0)} in the following way.
First, we find that $f(s)=s\tilde P(a,s|a)$ is a concave function $f''(s)\le 0$ within the interval $-\lambda_1\le s\le 0$, whereas  $g(s)\equiv s(s+\lambda_1)\tilde P(a,s|a)=(s+\lambda_1)f(s)$
is a convex function $g''(s)\ge 0$ in the same interval. Hence, we find that the tangent
$t_f(s)=\sigma_0+s\sigma_1$ to $f$ and the
tangent $t_g(s)=\lambda_1\sigma_0+(\sigma_0+\lambda_1\sigma_1)s$ to $g$ have roots
that sandwich $s=-\mu_1$ according to
\begin{equation}
\frac{\tilde\mu_1(a)}{1+\tilde\mu_1(a)/\lambda_1}=\frac{\lambda_1\sigma_0}{\sigma_0+\lambda_1\sigma_1}\le \mu_1(a)\le \frac{\sigma_0}{\sigma_1}=\tilde\mu_1(a),
\label{eq:sandwich}
\end{equation}
where the lower bound, $s=-\tilde\mu_1(1+\tilde\mu_1/\lambda_1)^{-1}$,
solves $t_g(s)=0$ and the upper bound, $s=-\tilde\mu_1$, solves $t_f(s)=0$.  
The chain of inequalities Eq.~\eqref{eq:sandwich} and its
implications, which we explore below, are the main result of this paper. Notably, in the limit of $U(a)\gg k_{\rm B}T$,
where $\tilde\mu_1 \to 0$ (i.e., $\tilde\mu_1\ll\lambda_1$ \cite{matk81}) holds,
the inequalities in Eq.~\eqref{eq:sandwich} saturate and provide an asymptotically exact value for $\mu_1$. 

We emphasize that the chain of inequalities \eqref{eq:sandwich}
holds for the slowest time-scale $\mu_1^{-1}$ of the first passage process as well
as for the slowest time-scale of the extremum functional, i.e. either the maximum or the minimum. 
For example, if $\mu_1^\text{min}(a)$ and $\mu_1^\text{max}(a)$
denote the large deviation limit of the minimum and maximum,
respectively, then the slowest first passage rate is given by $\mu_1(a)=\min[\mu_1^\text{max}(a),\mu_1^\text{min}(a)]$.
Since the maximum $\overline{m}_t=a\ge x_0$ after a long time $t$ will be more likely located at the ``right'' border of a confining potential, where $U(\overline{m}_t)\gg U(x_0)$, whereas the minimum $\underline{m}_t=a\le x_0$
will more likely move to the ``left'' border, where $U(\underline{m}_t)\gg U(x_0)$,
we will use Eq.~\eqref{eq:sandwich} to determine the minimum near the left boundary $a<x_0$
and to determine the maximum if $a>x_0$ is closer to the right
boundary. 

To be more specific, we use $\tilde \mu_1(a)\simeq\mu_1^\text{max}(a)$ self-consistently for the large deviation limit of the maximum,
whenever $\tilde \mu_1(a)\searrow$ is monotonically decreasing with
increasing $a\nearrow$, whereas we use $\tilde
\mu_1(a)\simeq\mu_1^\text{min}(a)$ for the large deviation limit of
the minimum, whenever $\tilde \mu_1(a)\searrow$ is monotonically
decreasing with decreasing $a\searrow$. Notably, if $a$ is located at
a reflecting boundary, where formally $U(x)=\infty$ for $x\le a$, we
immediately get $\mu_1(a)=\mu_1^\text{min}(a)$ and
$\mu_1^\text{max}(a)=\infty$, since for any $x_0>a$ the maximum
$\overline{m}_t$ cannot reach any value below $x_0$ and hence $a$
certainly cannot correspond to the maximum.\footnote{A similar finding can be found in \cite{hart18a_arxiv}, where $\mu_1(a)=\mu_1^\text{min}(a)$ corresponds to a first passage to $a$, entering from the right, and $\mu_1(a)=\mu_1^\text{max}(a)$ corresponds to a first passage to $a$, entering from the left. For example, if
a reflecting boundary is located at $a$ with $U(x)=\infty$ for $x\le
a$, it is impossible to enter $a$ from the left.}

\section{Examples}
\label{sec:examples}
\subsection{Statistics of  maxima in the Ornstein Uhlenbeck process}
\label{sec:OU}
\begin{figure}
 \centering
 \includegraphics{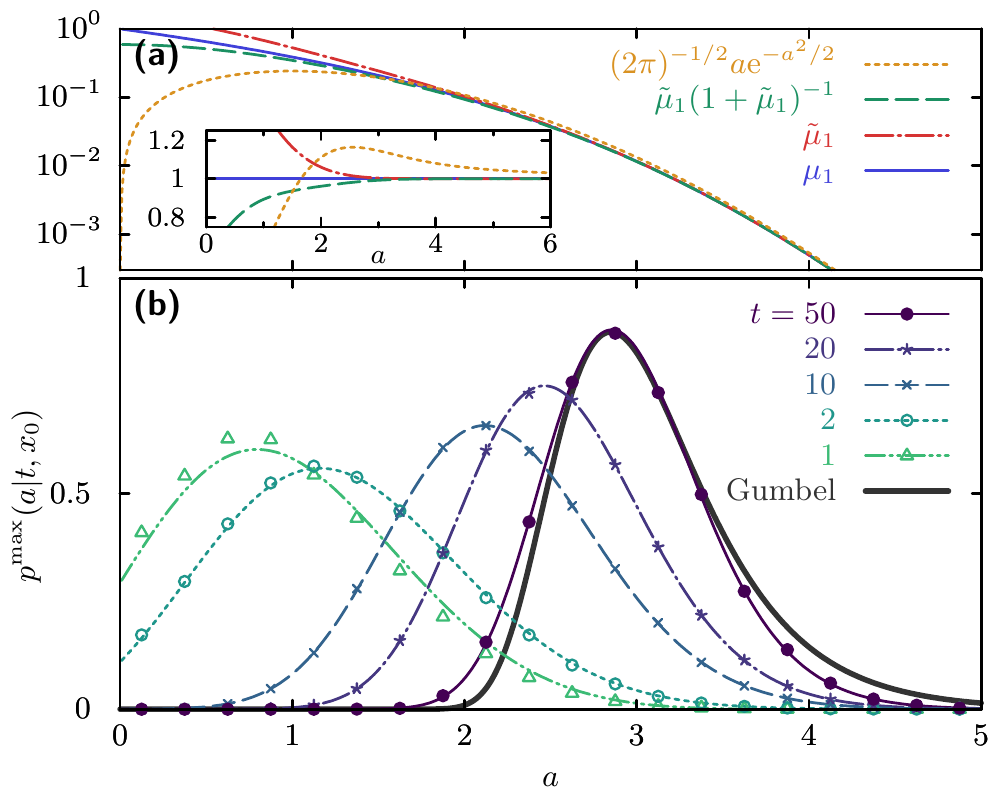} 
 \caption{Probability density of the maximum and its large deviation limit for the Ornstein-Uhlenbeck process.
 (a)~Large deviation eigenvalue $\mu_1(a)$ (solid blue line) compared to
   the approximation from Eq.~\eqref{eq:mu1a_OU} (dash-dotted red
   line) and the lower bound Eq.~\eqref{eq:sandwich} (dashed green line); the dotted line
 represents the asymptotic approximation  $(2\pi)^{-1/2}a\e^{-a^2/2}\simeq \mu_1(a)$
 from, e.g., Refs.~\cite{ricc88,maju14arxiv,gode16}. The inset shows
 the same result but scaled by $\mu_1$, which was determined numerically
 by the root, $s\tilde P(a,s|a)=0$, closest to the origin $s=-\mu_1$ (see, e.g., \cite{alil05}). (b)~Probability densities of the maximum $\pmax(a|t,x_0)$ (symbols)  are obtained from simulating  $10^5$ trajectories
 for each time $t=1,2,10,20,50$;
 the lines represent the large deviation limit
 $\pmaxLD(a|t,x_0)=\partial_a w_1(a,x_0)\e^{-\mu_1(a)t}$, where
 $\mu_1$ adopted from the upper panel~(a) and $w_1$ is determined from
 Eq.~\eqref{eq:w1_OU}. The thick gray line represents the Gumbel
 density
 $g(a,\eta,\gamma)=\gamma^{-1}\e^{-(a-\eta)/\gamma}\exp[-\e^{-(a-\eta)/\gamma}]$
 with arbitrarily chosen parameters $\eta=2.85$ and $\gamma=0.42$ \cite{gumb58}. The initial
 condition was $x_0=0$ and the symbols are obtained from Brownian
 dynamics simulations with a time increment $\dd t =10^{-5}$.
 }
 \label{fig:OU}
\end{figure}

As our first example we consider the Ornstein-Uhlenbeck process
with $U(x)=x^2/2$.
The corresponding propagator in the time domain is well known and reads \cite{gard04}
 \begin{equation}
 P(a,t|x_0)=\frac{1}{\sqrt{2\pi(1-\e^{-2t})}}\exp\bigg[-\frac{(a-x_0\e^{-t})^2}{2(1-\e^{-2t})}\bigg]
 \label{eq:propagator_OU}
\end{equation}
with a Gaussian equilibrium density $\Peq(a)=P(a,\infty|x_0)=(2\pi)^{-1/2}\exp(-a^2/2)$.
Inserting Eq.~\eqref{eq:propagator_OU} into Eq.~\eqref{eq:LD_approx}
yields for $a\ge0$ the following 
approximation for the large deviation eigenvalue of the density of the maximum\footnote{For $a\le 0$ Eq.~\eqref{eq:mu1a_OU} approximates the large deviation limit of the minimum functional.}
\begin{equation}
 \tilde\mu_1(a)=\int_0^\infty \bigg[\frac{1}{\sqrt{1-\e^{-2t}}}\exp\bigg(\frac{a^2\e^{-t}}{1+\e^{-t}}\bigg)-1\bigg]\dd t\simeq\mu_1(a).
 \label{eq:mu1a_OU}
\end{equation}
The relaxation eigenvalues are integers $\lambda_k=k$ with $k=0,1,\dots$, such that Eq.~\eqref{eq:sandwich} translates into
\begin{equation}
\tilde\mu_1(1+\tilde\mu_1)^{-1}\le\mu_1\le \tilde\mu_1,
\label{eq:sandwich_OU}
\end{equation}
where the upper limit $\tilde\mu_1$
is depicted in Fig.~\ref{fig:OU}a as the dash-dotted red line, the
lower limit $\tilde\mu_1(1+\tilde\mu_1)^{-1}$ as the dashed green
line, and the exact value $\mu_1$, determined as described below, is given by the solid
blue line. The inset displays the same results but scaled by the exact
value $\mu_1$. 
We emphasize that it is not necessary to determine $\mu_1$
in order to show that Eq.~\eqref{eq:mu1a_OU} asymptotically saturates
when $\tilde\mu_1$ approaches zero in the limit 
of large $a$, since $\tilde\mu_1\simeq\mu_1$ follows immediately from $\tilde\mu_1(1+\tilde\mu_1)^{-1}\simeq \tilde\mu_1$ (for $\tilde\mu_1\to0$)
as well as from Eq.~\eqref{eq:sandwich_OU}. 
For completeness,
we also present  in Fig.~\ref{fig:OU}a (dotted line) the long time asymptotics,
$\mu_1\simeq (2\pi)^{-1/2}a\e^{-a^2/2}$,
for the limit $a\to\infty$, which have been reported previously \cite{ricc88,maju14arxiv,gode16}.

In order to determine the large deviation eigenvalue $\mu_1$ and weight $w_1$
entering $\pmaxLD(a|t,x_0)=\partial_aw_1(a,x_0)\e^{-\mu_1(a)t}$,
we Laplace-transform the propagator \eqref{eq:propagator_OU} in time ($t\to s$),
which for $x_0\le a$ yields (see also \cite{sieg51}) 
\begin{equation}
\tilde P(a,s|x_0)=\Gamma(s)2^s \Peq(a)H_{-s}(-a/\sqrt{2})H_{-s}(x_0/\sqrt{2}),
\label{eq:propagator_OU_2}
\end{equation}
where  $\Gamma(s)$ is the complex gamma function and $H_s(y)$ is the generalized Hermite polynomial. Inserting Eq.~\eqref{eq:propagator_OU_2}
into the renewal theorem \eqref{eq:renewal_Laplace} yields \cite{sieg51,marg05,mart19}
\begin{equation}
 \tilde\wp_a(s|x_0)=\frac{H_{-s}(-x_0/\sqrt{2})}{H_{-s}(-a/\sqrt{2})},
\end{equation}
where $s=-\mu_1$ is the root, $H_{-s}(-a/\sqrt{2})=0$, closest to the origin
such that the weight in Eq.~\eqref{eq:w1} becomes \cite{ricc88,alil05}
\begin{equation}
 w_1(a)=-\frac{H_{\mu_1}(-x_0/\sqrt{2})}{\mu_1h_{\mu_1}(-a/\sqrt{2})},
 \label{eq:w1_OU}
\end{equation}
where we introduced $h_s(y)\equiv\partial_sH_s(y)$. For convenience we determined $\mu_1$ and $w_1$ numerically according to  Ref.~\cite{alil05}.%
\footnote{%
We note that with the eigenfunctions $\psi_k(a)=\Peq(a)(k!2^k)^{-1/2}H_k(a/\sqrt{2})$ (see, e.g., Ref.~\cite{gard04}) and Eqs.~\eqref{eq:P(a,s|x_0)} and  \eqref{eq:Taylor_series}
we can formally identify
\begin{equation}
 \sigma_0=\Peq(a)\qquad\text{and}\qquad\frac{\sigma_n}{n!}=\Peq(a)(-1)^{n+1}\sum_{k=1}^\infty\frac{H_k(a/\sqrt{2})^2}{k!2^kk^n},
 \nonumber
\end{equation}
which with Eq.~\eqref{eq:mu1_Newton} would be an alternative but
equivalent approach for determining $\mu_1$ as done, e.g. in Ref.~\cite{hart18a_arxiv}.
}
The results in Fig.~\ref{fig:OU}a confirm the validity of
the chain of inequalities in Eq.~\eqref{eq:sandwich}, which, as
already mentioned, become asymptotically
tight in the limit of high values of the potential, $U(a)=a^2/2\gg 1$.

The lines in Fig.~\ref{fig:OU}b
represent the large deviation limit of the density of the maximum
$\pmaxLD(a|x_0)=\partial_a w_1(a,x_0)\e^{-\mu_1(a)t}$, which agree
rather well with the density of the maximum $\pmax(a|x_0)$ (symbols) obtained from
Brownian dynamic simulations with a time step $\dd t=10^{-5}$ using $10^5$ trajectories.

We note that the large deviation limit $\pmaxLD$ (see lines in Fig.~\ref{fig:OU}b)
approximates the density of the maximum $\pmax$
quite well already on relatively short time-scales
$t\sim\lambda_1^{-1}=1$ (see triangles and dash-dotted light green
line), where $\lambda_1^{-1}$ represents the equilibration time of the
Ornstein-Uhlenbeck process. Notably, even for long times (see, e.g.,
$t=50$ in Fig.~\ref{fig:OU}b), the left and right tails of the density of the maximum remain asymmetric
yet still deviating from a Gumbel distribution \cite{gumb58} (see
thick gray line in Fig.~\ref{fig:OU}b). This indicates that the extreme value theorem for sequences of uncorrelated random
variables becomes valid on much longer time-scales.
Therefore, the large deviation limit presented here allows us to approximate extreme value statistics
exceptionally well despite the fact that the extreme value theorem does not yet apply.\footnote{
We find that the probability density of the  maximum approaches a Gumbel density on extremely large time-scales  $t\gtrsim10^3$. The underlying assumptions are
the approximation $\mu_1\simeq a(2\pi)^{-1/2}\e^{-a^2/2}$ (see inset of Fig.~\ref{fig:OU}a for deviations) and $w_1\simeq 1$ (which holds for $a\gtrsim3$).}

\subsection{Density of the minimum of the confined Bessel process}
\label{sec:Bessel}
\begin{figure}
 \centering
 \includegraphics{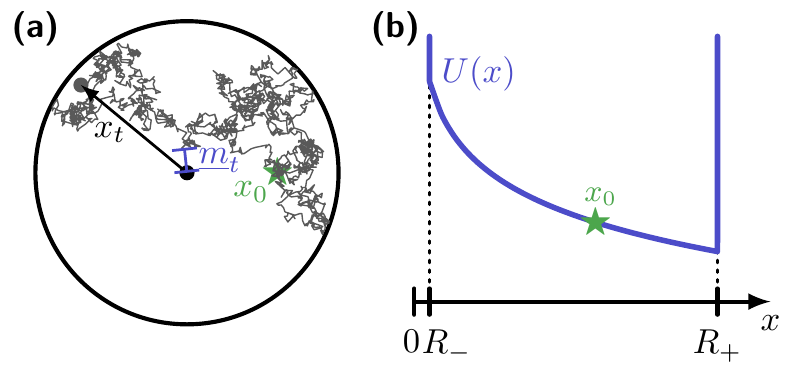}
 \caption{Graphical illustration of the Bessel process. (a)~Trajectory
   of a Brownian motion starting from $x_0$ and ending at distance $x_t$ after time $t$ in $d=2$ dimensions. The minimum of the distance is indicated as $\underline{m}_t$. (b)~
Effective potential $U(x)=-\ln(x^{d-1})$, where $x$ denotes the distance to the origin and
 $R_-$ the inner radius of the confinement and $R_+$ the outer radius of the volume.}
 \label{fig:ill_Bessel}
\end{figure}

In our second example we consider the minimal distance to the origin
of Brownian motion inside a $d$-dimensional sphere with inner radius
$R_-\ge0$ and a reflecting boundary at $R_+<\infty$ (see Fig.~\ref{fig:ill_Bessel}a for an illustration with $R_-=0$ and $d=2$).
The distance from the origin $x_t$ (i.e. the radius) at time $t$
within the interval $R_-\le x\le R_+$ obeys the Langevin equation
\begin{equation}
\dot x_t=\frac{d-1}{x_t}+\xi_t
\label{eq:Bessel_SDE}
\end{equation}
where $\avg{\xi_t}$ and $\avg{\xi_t\xi_{t'}}=2\delta(t-t')$. This process
is also known as the Bessel process \cite{pitm75,bark14}. We note that
the maximum excursion of the free Bessel process (see e.g. \cite{sche10}), which in the present context corresponds to the limiting case
with $R_+=\infty$ and will not be considered here, allows in the
specific case of $d=3$ a mapping onto a simpler problem for the
1-dimensional Brownian motion \cite{pitm75}.

Comparing Eq.~\eqref{eq:Langevin} and Eq.~\eqref{eq:Bessel_SDE} allows
us to identify the geometric free energy  $U(x)=-(d-1)\ln x$ of purely
entropic origin and accounts for the invariance with respect to
angular degrees of freedom $\propto x^{d-1}$ (see Fig.~\ref{fig:ill_Bessel}b). The equilibrium measure
corresponds to a uniform distribution in a $d$-dimensional
hyperspherical shell and is given by $\Peq(x)=dx^{d-1}/(R_+^d-R_-^d)$.

For simplicity we here from restrict our discussion to the case $d=3$,
yielding the Fokker-Planck equation
\begin{equation}
\frac{\partial}{\partial t} P(x,t|x_0)=\bigg[\frac{\partial^2}{\partial x^2}-\frac{\partial}{\partial x}\frac{2}{x}\bigg]P(x,t|x_0)
\label{eq:Bessel_FPE}
\end{equation}
with zero flux boundary condition $J(R_\pm,t|x_0)= 0$, where $J(x,t|x_0)\equiv (2/x-\partial_x)P(x,t|x_0)$.
We emphasize that the probability density $P$ is normalized according
to $\int_{R_-}^{R_+}P(x,t|x_0)\dd x=1$, whereas the radial density,
discussed for example in \cite{redn01}, would correspond to $P(x,t|x_0)/(4\pi x^2)$ instead.
A Laplace transform in $t$
yields 
\begin{equation}
 \bigg[\frac{\partial^2}{\partial x^2}-\frac{\partial}{\partial x}\frac{2}{x}-s\bigg]\tilde P(x,s|x_0)=-\delta(x-x_0),
 \label{eq:Bessel_FPE_inhom}
\end{equation}
where the solution $\tilde P(x,s|x_0)$ can be constructed from the two solutions of the
homogeneous problem, $v_1(x,s)=x\e^{-x\sqrt{s}}$ and
$v_2(x,s)=x\e^{x\sqrt{s}}/\sqrt{s}$ obtained by setting the right hand side of Eq.~\eqref{eq:Bessel_FPE_inhom} to zero. The Laplace transform
of the propagator for a Brownian particle confined between $R_-=a$ and
$R_+=R$ in turn
reads
\begin{align}
\tilde P(a,s|x_0)=
\frac{a^2\sinh \left[\sqrt{s} \left(R-x_0\right)\right]-a^2 R \sqrt{s} \cosh \left[\sqrt{s} \left(R-x_0\right)\right]}{x_0 (1-a R s) \sinh \left[\sqrt{s} (R-a)\right]-\sqrt{s} x_0 (R-a) \cosh \left[\sqrt{s} (R-a)\right]}.
\label{eq:Propagator_Bessel}
\end{align}
We are allowed to choose $R_-=a$, since the
first passage time distribution from $x_0$
to $a\le x_0$ is not affected by the potential $U(x)$ in the region $x\le a$,
where $R_-=a$ formally corresponds to $U(x)=\infty$ for $x\le a$ (see also the discussion at the end of Sec.~\ref{sec:max_vs_FP}). Most importantly, setting $R_-=a$
removes all roots of
$\tilde P(a,s|a)$, which would account for the maximum of the Bessel
process. In other words, in the presence of a reflecting boundary at
$a$ every single root
of $\tilde P(a,s|x_0{=}a)$ from Eq.~\eqref{eq:Propagator_Bessel} is
indeed a first passage time scale for approaching $a$ for the first
time from above (for more details on the influence of boundary
condition please see \cite{hart18a_arxiv}).
Moreover, the limit $R=R_+=\infty$, which is not considered here,
would allow us to map the 3d-Bessel process to 1d Brownian motion \cite{pitm75} with $\tilde \wp_a(s|x_0)=\tilde P(a,s|x_0)/\tilde P(a,s|a)\to (a/x_0)\exp[\sqrt{s}(a-x_0)]$,
which would in turn yield the Levy-Smirnov
density\footnote{The Levy-Smirnov
density is defined as $\wp_a(t|x_0)= (a/x_0)\times(x_0-a)/\sqrt{4\pi t^{3}}\times\e^{-(x_0-a)^2/(4t)}$.}. 

\begin{figure}\centering
\includegraphics{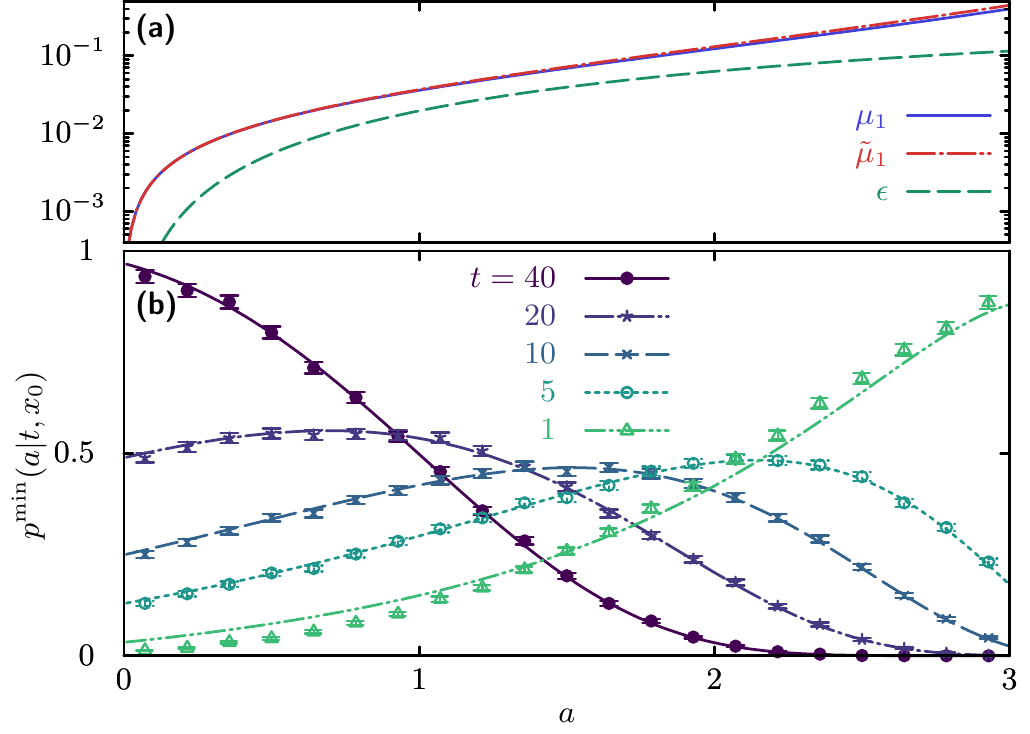}%
 \caption{Probability density of the minimum of a 3d Bessel process.
 (a) Slowest time scale $\mu_1$ (solid blue line) and
 its approximation $\tilde \mu_1$ (dash-dotted red line) given by
 Eq.~\eqref{eq:mu1a_Bessel} as a function function of the distance to
 the origin $a$;
 we truncated Eq.~\eqref{eq:mu1_Newton} after $n=10$ to calculate
 $\mu_1$. The corresponding relative deviation $\epsilon(a)=[\tilde
   \mu_1(a)- \mu_1(a)]/\mu_1(a)$ is depicted by the dashed green line.
 (b)~
 Probability density of the minimum (symbols), $\pmin(a|t,x_0)$,
 sampled from $10^5$ Brownian trajectories obtained by evolving
 Eq.~\eqref{eq:Bessel_SDE} with time increment $\dd t=10^{-5}$ and using $R_-=0$;
error-bars indicate 95\,\% confidence intervals.
The large deviation limit $\pminLD(a|t,x_0)=-\partial_aw_1(a,x_0)\e^{-\mu_1(a)t}$ (lines) determined using the eigenvalues $\mu_1(a)$ from (a) and the weight $w_1$
 from Eqs.~\eqref{eq:w1} and ~\eqref{eq:Propagator_Bessel}.
All results correspond to an Initial distance $x_0=3$ and an outer radius $R_+=5$.}%
 \label{fig:Bessel}%
\end{figure}

For $R<\infty$, we use  Eq.~\eqref{eq:Propagator_Bessel} to identify
the Taylor coefficients of $s\tilde P(a,s|a)$, which we denote by
$\sigma_n$ according to Eq.~\eqref{eq:Taylor_series}. The exact
smallest eigenvalue $\mu_1$ is then determined using Eq.~\eqref{eq:mu1_Newton}. The results are presented in Fig.~\ref{fig:Bessel}a (see solid blue line), where we also compare $\mu_1$
to the approximation $\tilde\mu_1$ from Eq.~\eqref{eq:LD_approx} (see dash-dotted red line),
which for the 3$d$-Bessel process reads%
\begin{equation}
 \tilde\mu_1(a)=\frac{\sigma_0}{\sigma_1}=
\frac{15 a \left(a^2+a R+R^2\right)}{(a-R)^2 \left(a^3+3 a^2 R+6 a R^2+5 R^3\right)}\simeq \mu_1(a).
\label{eq:mu1a_Bessel}
\end{equation}
Eq.~\eqref{eq:mu1a_Bessel} delivers the exact value for $\mu_1$
in the limit $a\to 0$ as shown in Fig.~\ref{fig:Bessel}a, where the
relative deviation $\epsilon\equiv (\tilde{\mu}_1-\mu_1)/\mu_1$
vanishes in the limit $a\to0$ (see dashed green
line).
It should be noted that $\mu_1$,
given by the series Eq.~\eqref{eq:mu1_Newton}, is in fact an explicit solution
of the transcendental equation $R\sqrt{\mu_1}=\tan[(R-a)\sqrt{\mu_1}]$
in the form of a Newton's series.

To rationalize why Eq.~\eqref{eq:mu1a_Bessel} becomes asymptotically
exact as $a$ approaches zero, we recall that $\lambda_1\approx0.81$ is
the slowest relaxation rate corresponding to $R_-=0$, which solves
$R\sqrt{\lambda_1}=\tan(R\sqrt{\lambda_1})$ (here using $R=5$). Since
Eq.~\eqref{eq:Propagator_Bessel} obeys a reflecting boundary
condition at $a$, we have that $\lambda_1(a)\ge\lambda_1\approx 0.81$, i.e.
the eigenvalue $\mu_1(a)$ is bounded by
$\tilde\mu_1(1-\tilde\mu_1/0.81)\le\mu_1\le\tilde\mu_1$. Hence for
asymptotically high potentials (here $U(a)=-2\ln a\to \infty$ as $a\to
0$) the inequality Eq.~\eqref{eq:sandwich} renders $\tilde \mu_1$
asymptotically exact as soon as $\tilde\mu_1 (a)/0.81\to 0$.

In Fig.~\ref{fig:Bessel}b we compare the density of the minimum $\pmin$ (symbols),
obtained from simulations of $10^5$ trajectories (with $R_-=0$,
$R_+=R=5$ and starting condition $x_0=3$), to the corresponding large
deviation limit $\pminLD(a|t,x_0)=-\partial_a
w_1(a,x_0)\e^{-\mu_1(a)t}$ (lines), where we determined $w_1$ using
Eq.~\eqref{eq:w1} and took the exact $\mu_1(a)$ obtained using Eq.~\eqref{eq:Propagator_Bessel}. The error bars in the simulation results denote 95\,\% confidence intervals.

By design the large deviation limit $\pminLD$ approaches the density of the minimum $\pmin$ in the long time limit,
which is perfectly corroborated by simulation results for $t=40\,(\gg
\lambda_1^{-1}\approx1.2)$ in Fig.~\ref{fig:Bessel}b. Notably, $\pminLD$
(see solid dark-blue line in Fig.~\ref{fig:Bessel}b) approximates quite well the full probability density of the minimum $\pmin$ (see filled circles).

To our surprise, the large deviation limit $\pminLD$ can approximate $\pmin$ even for smaller times (e.g., $t=1$), i.e. those that are shorter than the equilibration time $\lambda_1^{-1}\approx1.2$ (cf. dash-dotted green line vs. open triangles),
which can be explained as follows.
For any $a$ within  $0\le a\le 3=x_0$
we find a spectral gap $\mu_1(a)\ll \mu_2(a)$, which for $1.5\lesssim a\le 3$
also satisfies $\mu_2(a)\gg\lambda_1 (a)\approx0.81$. This in turn implies that for $t=1\sim\lambda_1^{-1}$ the condition
$\e^{-\mu_1t}\gg \e^{-\mu_2t}$ is still met, whereas the relative deviations between $\pmin$
and $\pminLD$ (i.e., between the open triangles and the dash-dotted green line)
become substantial for small values of $a$ (see $a\le 1$) and $\mu_2(a)$
concurrently approaches $\lambda_1\simeq0.8$. 
Once the time exceeds $\lambda_1 t\approx0.81t\gg1$, the condition $\e^{-\mu_1t}\gg
\e^{-\mu_2t}$ is satisfied for any value of $a$, and thus $\pminLD$
approximates $\pmin$ over the full range (see symbols and lines in
Fig.~\ref{fig:Bessel}b for $t\ge5$). Therefore, $\pminLD$ approximates
$\pmin$ rather well for any value $a$ and on all time scales longer than the equilibration time scale ($t\gg \lambda_1^{-1}$).

Let us finally discuss the ``ultimate'' long time
limit ($t\to\infty$) in which the density of the minimum $\pmin(a|t,x_0)$ will be sharply peaked around the shortest distance $a=0$. Inspecting Eq.~\eqref{eq:mu1a_Bessel}
one can easily find $\mu_1(a)=3 a/R^3+\mathrm{O}(a)$. Moreover, at  high values of the potential the weight
becomes
 $w_1(a,x_0)=1+\mathrm{O}(a)$ implying that the limiting density
becomes $\pmin(a|t,x_0)\to 3tR^{-3}\e^{-3at/R^3}$, which is an
exponential distribution falling into the class of Weibull distributions. At $t=40$ (see Fig. \ref{fig:Bessel}b) the density of the minimum
 still
 qualitatively deviates from an exponential density;
 while the exponential density is a convex function of $a$
 the resulting curve from Fig. \ref{fig:Bessel}b for $t=40$
clearly did not yet reach a convex shape in $a$. While the Ornstein-Uhlenbeck process shows a Gumbel distribution in the limit $t\to\infty$, the Bessel process 
provides an example in which the extreme value distribution
falls into the class of Weibull distributions.

\section{Concluding perspectives}
\label{sec:conclusion}
We used the link between first passage and extremum functionals of
reversible ergodic Markov processes in order to formulate the probability density of extreme values in
terms of the first passage times. We pushed the
connection between these two functionals even further, by
utilizing the duality between first passage and relaxation processes
\cite{hart18,hart18a_arxiv}, which allowed us to determine the
statistics of extremes from transition probability densities
describing the relaxation towards equilibrium. In their present form
our results hold for diffusion in effectively one-dimensional
potential landscapes that are sufficiently confining to allow for a
discrete eigenspectrum of the corresponding Fokker-Planck
operator. Our findings provide a new and deeper perspective on the
study of extrema of asymmetric diffusion processes beyond a constant drift.
We emphasize that the full probability density of extreme values
($\pmax$ or $\pmin$) on arbitrary time-scales still
requires the knowledge of the eigenspectrum of the Fokker-Planck operator.

To avoid an eigendecomposition of the Fokker-Planck operator entirely,
we established the long time asymptotics of the distribution of extreme values, $\pmaxLD\simeq\pmaxLD$ (or $\pminLD\simeq\pmin$),
which accounts for the slowest decaying mode $\propto\e^{-\mu_1t}$ ignoring all faster decaying contributions ($\propto\e^{-\mu_2 t},\e^{-\mu_3 t},$ etc.).
In this large deviation limit we determined explicit bounds on 
the exact slowest time-scale $\mu_1^{-1}$, and showed that these asymptotically tightly bound $\mu_1$ from above and from below, which is the central result of this paper.

We illustrated the usefulness of our results by analyzing the statistics of maximum value of the Ornstein-Uhlenbeck process and
the minimal distance to the origin of a confined 3d Brownian motion (Bessel process).
Our examples underline that the large deviation limit, albeit designed
to be asymptotically exact for infinitely long times, approximates the
density of the maximum surprisingly well even on relatively short
times comparable to the relaxation time, $t\gtrsim1/\lambda_1$. Since
$t=1/\lambda_1$ reflects the time-scale on which the process
tends to decorrelate from the initial condition, the present results
describe the statistics of extrema in presence of weak but
non-vanishing correlations, and hence go beyond the three classes of
limit laws for non-correlated random variables, i.e. the Gumbel,
Fr\'echet, and Weibull
distributions~\cite{gumb58,bert06,sabh07,krug07,luca12,werg13} as
demonstrated on hand of the Ornstein-Uhlenbeck and Bessel
process. More generally, it would be interesting to systematically investigate the effect of the potential shape on the limiting extreme value distribution as in Ref.~\cite{sabh07}.

The remarkable accuracy of the approximation can readily be explained by the interlacing of first passage and relaxation time-scales ($\mu_1\le\lambda_1\le\mu_2\ldots$) \cite{hart18,hart18a_arxiv}, which renders all higher contributions ($\propto\e^{-\mu_2 t},\e^{-\mu_3 t}\ldots$) negligibly small compared to the large deviation limit $\propto \e^{-\mu_1t}$ once the condition $t\gtrsim\lambda_1^{-1}$ is met. 

Our results can be extended and generalized in various
ways. Extending the formalism presented here to systems obeying a discrete state Master equation would be straightforward; for example, our main result Eq.~\eqref{eq:sandwich}
would still hold by formally replacing the probability densities from Eq.~\eqref{eq:LD_approx_int}
by the corresponding state probabilities~\cite{hart18a_arxiv}. Interesting and challenging extensions could include the consideration of
trapping times \cite{cond07,krue14}, spatial disorder \cite{gode16a} and multi-channel transport \cite{gode17}, as well as extreme value statistics in discrete-state Markov processes with a broken time-reversal symmetry \cite{neri17,sing17arxiv}.

\ack

The financial support from the German Research
Foundation (DFG) through the Emmy Noether Program
``GO 2762/1-1'' (to AG) is gratefully acknowledged.

\section*{References}

\providecommand{\href}[2]{#2}\begingroup\raggedright\endgroup

\end{document}